\documentclass[11pt,english]{revtex4}
\usepackage{lmodern}

\usepackage[T1]{fontenc}
\usepackage[latin9]{inputenc}
\usepackage{amsmath}
\usepackage{amssymb}
\usepackage{esint}

\makeatletter
\@ifundefined{textcolor}{}
{%
 \definecolor{BLACK}{gray}{0}
 \definecolor{WHITE}{gray}{1}
 \definecolor{RED}{rgb}{1,0,0}
 \definecolor{GREEN}{rgb}{0,1,0}
 \definecolor{BLUE}{rgb}{0,0,1}
 \definecolor{CYAN}{cmyk}{1,0,0,0}
 \definecolor{MAGENTA}{cmyk}{0,1,0,0}
 \definecolor{YELLOW}{cmyk}{0,0,1,0}
 }

\usepackage{latexsym}\usepackage{bm}

\makeatother

\makeatother

\usepackage{babel}

\makeatother

\usepackage{babel}

\begin{document}

\title{BORN-INFELD EXTENSION OF NEW MASSIVE GRAVITY}

\author{\.{I}brahim Güllü }

\email{e075555@metu.edu.tr}

\affiliation{Department of Physics,\\
 Middle East Technical University, 06531, Ankara, Turkey}

\author{Tahsin Ça\u{g}r\i{} \c{S}i\c{s}man}

\email{sisman@metu.edu.tr}

\affiliation{Department of Physics,\\
 Middle East Technical University, 06531, Ankara, Turkey}

\author{Bayram Tekin}

\email{btekin@metu.edu.tr}

\affiliation{Department of Physics,\\
 Middle East Technical University, 06531, Ankara, Turkey}

\date{\today}
\begin{abstract}
We present a three-dimensional gravitational Born-Infeld theory which
reduces to the recently found new massive gravity (NMG) at the quadratic
level in the small curvature expansion and at the cubic order reproduces
the deformation of NMG obtained from AdS/CFT. Our action provides
a remarkable extension of NMG to all orders in the curvature and might
define a consistent quantum gravity. 
\end{abstract}

\pacs{04.60.Kz,04.50.-h,04.60.-m}

\maketitle
Mass of the graviton is a subtle issue: no satisfactory Higgs-type
mechanism seems to exist for spin-2 particles. Therefore, one usually
supplies a Pauli-Fierz type hard mass to the graviton, which is quadratic
in the fluctuations of the metric around a fixed (maximally symmetric)
background with the unique ghost-free combination $m^{2}\left(h_{\mu\nu}^{2}-h^{2}\right)$.
Pauli-Fierz mass comes with a price: general covariance is lost, ghosts
appear beyond the tree level and in the interacting theory \cite{boulware},
and $m^{2}=0$ theory is discretely disconnected from $m^{2}\rightarrow0$
theory. To solve these problems, one must search for other non-Pauli-Fierz
type masses. At least in three dimensions, the theory (NMG) recently
introduced by Bergshoeff \emph{et al} \cite{bht} provides a nonlinear,
generally covariant extension of the Pauli-Fierz massive gravity (for
the mostly plus signature): \begin{equation}
I_{\text{NMG}}=\frac{1}{\kappa^{2}}\int d^{3}x\,\sqrt{-\det g}\left[-R+\frac{1}{m^{2}}\left(R_{\mu\nu}^{2}-\frac{3}{8}R^{2}\right)\right].\label{eq:NMG_action}\end{equation}
 This action (or its cosmological extension) defines a massive spin-2
particle around Minkowski (or (anti)-de-Sitter) background. Tree-level
ghost freedom, Newtonian limits, classical solutions, supergravity
extensions, etc have been worked out in \cite{bht,deser,nakasone,clement,gullu1,gullu2,giribet,gurses,troncoso,liu-sun,sezgin}.
NMG, being the only ghost-free super-renormalizable theory, since
its four-dimensional cousin is renormalizable \cite{stelle}, may
prove to be a useful laboratory for quantum gravity. Therefore, its
possible extensions are invaluable. Here, we present extensions of
NMG in terms of determinantal Born-Infeld actions which have appeared
in various contexts in the past as means of generalizing Einstein's
gravity theory, Maxwell's theory, Yang-Mills theory, and found their
natural place in string theory as D-brane actions. {[}For details
see \cite{gibbons}.{]} In the gravitational context, Deser and Gibbons
\cite{gibbonsDeser} studied conditions on viable Einstein Born-Infeld
actions in four dimensions. One such condition is ghost freedom which
necessarily removes all the quadratic terms after small curvature
expansion. On the other hand, in three dimensions, as we have seen
in the NMG case, since a proper combination of quadratic terms yields
ghost-free action, one can define a remarkably simple gravitational
Born-Infeld action that extends NMG. Below, we present the bare essentials
of the theory leaving the details for another work. Let us start with
the simplest case where there is no cosmological constant. The action\begin{equation}
I_{\text{BI}}=-\frac{4m^{2}}{\kappa^{2}}\int d^{3}x\,\left[\sqrt{-\det\left(-\frac{1}{m^{2}}\mathcal{G}\right)}-\sqrt{-\det g}\right],\label{eq:Born-Infeld}\end{equation}
 with $\mathcal{G}_{\mu\nu}\equiv R_{\mu\nu}-\frac{1}{2}g_{\mu\nu}R-m^{2}g_{\mu\nu}$
reduces to (\ref{eq:NMG_action}) upon use of the small curvature
expansion \begin{align}
\left[\det\left(1+A\right)\right]^{1/2} & =1+\frac{1}{2}\text{Tr}A+\frac{1}{8}\left(\text{Tr}A\right)^{2}-\frac{1}{4}\text{Tr}\left(A^{2}\right)\nonumber \\
 & \phantom{=}+\frac{1}{6}\text{Tr}\left(A^{3}\right)-\frac{1}{8}\text{Tr}\left(A^{2}\right)\text{Tr}A+\frac{1}{48}\left(\text{Tr}A\right)^{3}+O\left(A^{4}\right),\label{eq:det_exp}\end{align}
 up to order $O\left(A^{3}\right)$ %
\footnote{Another extension of NMG is realized with the action \[
I=-\frac{4m^{2}}{\kappa^{2}}\int d^{3}x\,\left\{ \sqrt{-\det\left[g_{\mu\nu}+\frac{1}{m^{2}}\left(R_{\mu\nu}-\frac{1}{6}g_{\mu\nu}R\right)\right]}-\sqrt{-\det\left(g_{\mu\nu}\right)}\right\} .\]
}. {[}We kept $O\left(A^{3}\right)$ terms in the above expansion for
later use below.{]} Note that to reproduce the Einstein-Hilbert action
at the first order, one necessarily uses the cosmological Einstein
tensor $\mathcal{G}_{\mu\nu}$. In fact, to be able to make a small
curvature expansion, a $g_{\mu\nu}$ is needed in the first determinant.
In (\ref{eq:Born-Infeld}), $\sqrt{-\det g}$ removes a \emph{fixed}
cosmological constant coming from the leading-order expansion. To
accommodate a general cosmological constant and reproduce the cosmological
NMG (CNMG) theory \cite{bht}\[
I_{\text{CNMG}}=\frac{1}{\kappa^{2}}\int d^{3}x\,\sqrt{-\det g}\left[-\left(R-2\Lambda\right)+\frac{1}{m^{2}}\left(R_{\mu\nu}^{2}-\frac{3}{8}R^{2}\right)\right],\]
 one can simply modify (\ref{eq:Born-Infeld}) to be \begin{equation}
I_{\text{CBI}}=-\frac{4m^{2}}{\kappa^{2}}\int d^{3}x\,\left[\sqrt{-\det\left(-\frac{1}{m^{2}}\mathcal{G}\right)}-\left(\frac{\Lambda}{2m^{2}}+1\right)\sqrt{-\det g}\right],\label{eq:CBI_action}\end{equation}
 where $\mathcal{G}$ is given exactly as above. Note that to conform
with the second reference in \cite{bht}, one could set $\Lambda=m^{2}\lambda$
where $\lambda$ is a dimensionless number. In summary, (\ref{eq:CBI_action})
defines our minimally Born-Infeld-extended NMG for any (including
zero) cosmological constant. What is remarkable about this expression
is that only the cosmological Einstein tensor appears in the determinant.
It reproduces the only unitary theory NMG at the quadratic level,
and as we shall see below it also reproduces at the cubic level the
deformation of NMG obtained by the requirement that a holographic
$c$-theorem exists \cite{sinha}. {[}More recently, $O\left(R^{4}\right)$
matching was shown in \cite{gullu4}.{]}

At $O\left(R^{3}\right)$, let us compute what possible terms are
generated by the Born-Infeld action (\ref{eq:CBI_action}). At this
order using (\ref{eq:det_exp}) and defining $\mathcal{G}_{\mu\nu}\equiv G_{\mu\nu}-m^{2}g_{\mu\nu}$,
after a straightforward computation, one obtains\begin{align*}
O\left(R^{3}\right):\quad & -\frac{1}{m^{6}}\left[\frac{1}{6}\text{Tr}\left(g^{-1}Gg^{-1}Gg^{-1}G\right)-\frac{1}{8}\text{Tr}\left(g^{-1}Gg^{-1}G\right)\text{Tr}\left(g^{-1}G\right)+\frac{1}{48}\left[\text{Tr}\left(g^{-1}G\right)\right]^{3}\right]\\
 & =-\frac{1}{6m^{6}}\left[G^{\mu\nu}G_{\nu\alpha}G_{\phantom{\beta}\mu}^{\alpha}-\frac{3}{4}G_{\mu\nu}^{2}G_{\phantom{\alpha}\alpha}^{\alpha}+\frac{1}{8}\left(G_{\phantom{\alpha}\alpha}^{\alpha}\right)^{3}\right],\end{align*}
 where $G_{\phantom{\alpha}\alpha}^{\alpha}=-\frac{R}{2}$, $G_{\mu\nu}^{2}=R_{\mu\nu}^{2}-\frac{1}{4}R^{2}$
and $G^{\mu\nu}G_{\nu\alpha}G_{\phantom{\beta}\mu}^{\alpha}=R^{\mu\nu}R_{\nu\alpha}R_{\phantom{\alpha}\mu}^{\alpha}-\frac{3}{2}RR_{\mu\nu}^{2}+\frac{3}{8}R^{3}$.
Collecting all the terms together, we have\begin{align}
I_{\text{NMG-ext}} & =\frac{1}{\kappa^{2}}\int d^{3}x\,\sqrt{-\det g}\left[-\left(R-2\Lambda\right)+\frac{1}{m^{2}}\left(R_{\mu\nu}^{2}-\frac{3}{8}R^{2}\right)\right.\nonumber \\
 & \phantom{=\frac{1}{\kappa^{2}}\int d^{3}x\,\sqrt{-\det g}}\left.+\frac{2}{3m^{4}}\left(R^{\mu\nu}R_{\nu}^{\phantom{\nu}\alpha}R_{\alpha\mu}-\frac{9}{8}RR_{\mu\nu}^{2}+\frac{17}{64}R^{3}\right)+O\left(R^{4}\right)\right],\label{eq:NMG-ext}\end{align}
 which matches the result of Sinha \cite{sinha} obtained with the
help of AdS/CFT and the existence of a $c$-theorem in $\left(1+1\right)$-dimensional
CFT. This matching is highly non-trivial and lends support that (\ref{eq:CBI_action})
might define a consistent quantum gravity.

Now, let us discuss the non-minimal extensions. Using just the curvature,
and not its derivatives, and staying at the quadratic level one can
define the most general non-minimal Born-Infeld extension of the CNMG
in the following way: \begin{align*}
I_{\text{BI-non-minimal}} & =-\frac{1}{b}\int d^{3}x\,\sqrt{-\det\left(g+X\right)},\end{align*}
 where $X_{\mu\nu}\equiv a\left(R_{\mu\nu}+cg_{\mu\nu}R\right)+d\left(R_{\mu\alpha}R_{\phantom{\alpha}\nu}^{\alpha}+eg_{\mu\nu}R_{\alpha\beta}^{2}+fR_{\mu\nu}R+lg_{\mu\nu}R^{2}\right)$.
Observe that we have dropped the $\sqrt{-\det g}$ term. {[}Adding
some derivative terms here gives a quite interesting result \cite{gullu3}{]}.
Working out the expansions at the quadratic level, one has various
choices in trying to match the dimensional constants the $\kappa^{2}$,
$m^{2}$, $\Lambda$ and the dimensionless ratio $-\frac{3}{8}$ of
CNMG with the dimensional constants $a$, $b$, $d$ and the dimensionless
constants; $c$, $e$, $f$, and $l$ of the non-minimal Born-Infeld
theory. Here, for the sake of simplicity, we make our choice to be\begin{align}
I_{\text{BI-ext}} & =-\frac{1}{b}\int d^{3}x\,\sqrt{-\det\left[g_{\mu\nu}+a\left(R_{\mu\nu}+cg_{\mu\nu}R\right)+dR_{\mu\alpha}R_{\phantom{\alpha}\nu}^{\alpha}\right]},\label{eq:BI_non-min_simple}\end{align}
 where $a$, $b$, $c$ and $d$ are\[
a^{2}=-\frac{1}{2\Lambda}\left(\frac{9}{2m^{2}}+\frac{1}{\Lambda}\right),\qquad b=-\frac{\kappa^{2}}{2\Lambda},\qquad c=-\frac{1}{3}\left(1+\frac{1}{a\Lambda}\right),\qquad d=\frac{a^{2}}{2}+\frac{1}{\Lambda m^{2}}.\]
 So, (\ref{eq:BI_non-min_simple}) is also an extension of CNMG, perhaps
not as elegant as (\ref{eq:CBI_action}) and with the constraint $\Lambda<-\frac{2m^{2}}{9}$.
Therefore, with such an extension de-Sitter background is ruled out.
Of course, one may remove this constraint by considering the other
quadratic terms in the $X_{\mu\nu}$ tensor.

In conclusion, we have found the minimal (\ref{eq:CBI_action}) and
a non-minimal (\ref{eq:BI_non-min_simple}) Born-Infeld extensions
of the (cosmological) new massive gravity in three dimensions. Having
more powers of curvature, these theories are much better behaved in
the ultraviolet. In fact, at $O\left(R^{3}\right)$ our theory coincides
with the one obtained quite non-trivially from the AdS/CFT correspondence
with the requirement that a $c$-theorem exists \cite{sinha}. Further
work with our action in the context of AdS/CFT is needed. It would
also be interesting to see if these theories and their supersymmetric
generalizations can be derived from string theory D-brane actions.
In this communication, we have presented the basics of the model.
Elsewhere, we will return to discussions of classical solutions such
as BIonic solitons without singularities {[}see \cite{fiorini} for
a related discussion on curing singularities with the BI-type actions{]},
linearization of the theory around (anti)-de-Sitter spacetime and
the effects of cubic terms (\ref{eq:NMG-ext}) on the mass of the
graviton and more importantly on the unitarity of the theory. Around
the flat space, there is no problem about the unitarity at the tree
level. One can also couple matter fields to the metric: for example,
spin-1 Abelian gauge fields can simply be introduced with the substitution
$\det\left(g-\frac{1}{m^{2}}G\right)\rightarrow\det\left(g-\frac{1}{m^{2}}G+\alpha^{\prime}F\right)$.

\section{\label{ackno} Acknowledgments}

IG and BT are partially supported by T{Ü}B\.{I}TAK Kariyer Grant
104T177. TÇ\c{S} is supported by T{Ü}B\.{I}TAK PhD Scholarship.
We thank M Gürses and A Sinha, for bringing \cite{sinha} to our attention,
after we submitted the manuscript to the arXiv.

\end{document}